\documentstyle[psfig]{l-aa}
\begin{document}
   \thesaurus{
              08.16.6;
               02.18.5;
               02.19.2
               }
\title{An inverse Compton scattering (ICS) model of pulsar emission}
\subtitle{I. Core and conal emission beams}

\author{G.J. Qiao\inst{1},
W.P. Lin\inst{1,2}}
\institute{
Department of Geophysics,Peking University,Beijing 100871,PRC.
\and Beijig Astronomical Observatory,Chinese Academy of Sciences,Beijng 100080,PRC.
}
\offprints{W.P. Lin}
\date{Received 18.7.1996; accepted 12.8.97}
\maketitle

\begin{abstract}
Core and conal emission beams of radio pulsars have been identified
observationally (Rankin 1983,1993; Lyne and Manchester 1988), In the inner
gap model (Ruderman and Sutherland 1975, hereafter RS75), the gap continually breaks
down (sparking) by forming electron-positron pairs on a time scale of a few
microseconds (RS75). This makes a large amplitude low frequency wave to be produced
and which would be scattered by relativistic particles moving out from the
gap. Under this assumption, Qiao (1988a,1992) presented an Inverse Compton
Scattering (ICS) model for both core and conal emissions. This paper
presented a development of the model. Retardation and aberration effects due
to different radiation components emitted from different heights are
considered. The luminosity of pulsar radio
emission via the ICS process is discussed. Coherent emission by bunches of
particles is adopted and which is adequate to explain pulsar radiation.
The theoretical results are in agreement with observations very well.


\end{abstract}

\section{Introduction}

It's convincing that the emission beams of a radio pulsar can be divided
into two (core, inner conal) or three (plus an outer conal) emission
components through careful studies of the observed pulse profiles and
polarization characteristics (Rankin 1983a,1983b,1986,1990,1993; Lyne \&
Manchester 1988). Many pulsar profiles at meter wavelength are dominated by
core components. In usual polar cap models of pulsars, it is difficult to
get a central or ``core'' emission beam. Many current theoretical models can
only get a hollow cone emission beams. Thus, it is needed to make an effort
to get the core emission theoretically. Several authors such as Beskin et
al. (1988), Qiao (1988a,b), Wang et al. (1989) presented some models for the
core emission beam.

If the binding energy per ion in the neutron star surface is as large as 10
kev then ions will not be released, a pole magnetospheric vacuum gap (inner
gap) is formed (RS75). More
accurate variational calculations (e.g. Hillebrandt \& M\"uller 1976; Flowers
et al. 1977; K\"ossl et al. 1988) have revised downward the binding energy
to a few $kev$. According to RS75, primary particles are accelerated to
sufficient energies in the gap and emit high energy $\gamma$-quanta via
curvature radiation (hereafter CR), which in turn, ignite pair production
cascade to short out the acceleration potential (sparking). This is to say in
the calculations for the cascade of the inner gap only CR process is taking
into account. If the Inverse Compton Scattering (ICS) process in strong
magnetic fields is taking into account the potential drops across the gap
will be down, the ``binding energy difficulty'' will be released and the 
inner gap model (RS model) can still be sound, even if the binding energy
downward to a few $kev$ (Zhang and Qiao 1996, here after ZQ96; Qiao and Zhang
1996, here after QZ96; Zhang 1996; Zhang et al.1997).
The gap continually breaks down (sparking) by forming
electron-positron pairs on a time scale of a few microseconds (RS75),
this is inconsistent with the observed short-time scale structure (Hankins,1992).
A very large amplitude low frequency wave would be associated with the sparking,
which would be scattered by relativistic particles moving out (if the low
frequency wave can propagate in the magnetosphere of the neutron star, we
will discuss this a little below). Under the assumption that the observed
radio emission are produced by the ICS process of the high energy
particles (secondary) off the low frequency waves, we can get both core and
two conal components (Qiao 1988a,1992). The different emission components
are emitted from different heights (Qiao et al. 1992, Lin and Qiao 94, hereafter LQ96, and this paper, see bellow).

In this paper we present a development of the work of Qiao (1988a,b;1992)
which presented basic ideas and calculations for both core and two conal
emission beams. Both observations (Rankin 1993) and the theory (Qiao, et al.,
1992) show that the ``core'' emission is emitted at a place relatively close
to the stellar surface, the ``inner cone'' is emitted at a lower height, and
the ``outer cone'' is emitted at some greater height. If the different
emission components are emitted at different heights, two effects,
aberration and retardation effects, should be considered in the
calculations. Our result shows that these effects move the apparent
positions of the emission beams and change their shapes to be asymmetric to
the magnetic axis. McCulloch (1992) examined about 20 triple profiles and
found no example with the central component close to the leading component
but many close to the trailing component, our calculations fit the result
well. The basic idea, assumptions and the results of the calculations are
presented in section 2; some theoretical respects and observational facts
are discussed in sections 3.

This is one of a series papers about radio pulsar emission, on the
polarizations both linear and circular, the behavior of pulse profiles in
different frequencies, $\gamma$-ray emission and so on will be done later.

\section{An inverse Compton scattering (ICS) model}

\subsection{Assumption}

Along the line of RS inner gap model (RS75), Qiao (1988a,1992) presented a
model for radio emission of pulsars. The basic assumptions of the model are
as follows:

1. Neutron stars have dipole magnetic fields.

2. The Radio emission observed of pulsars is produced in an ICS process: a
low frequency wave (with angular frequency $\omega_0 \sim 10^6 s^{-1}$) is
scattered by high energy particles (with Lorentz factor $\gamma \sim
10^2-10^4$, this is the energy of the secondaries, see RS75 and if the ICS 
process of high energy particles off the thermal photon in or above the 
polar gap is taking into account, see Xia et al.1996, Bednarek et al. 1992,
ZQ96, Luo 1996, Zhang et al. 1997). 
The low frequency wave is produced in the inner gap
sparking (see RS75, the gap continually breaks down on a time scale of a few
microseconds, the angular frequency is corresponding to $\omega_0 \sim
10^6s^{-1}$, so the angular frequency $\omega_0 \sim 10^6 s^{-1}$ of the low
frequency wave is taking in the calculation below). The high energy
particles are the secondary particles produced near the gap (ZQ96
for self-consistent gap considering ICS-induced $\gamma-B$ process,
also see Zhang et al. 1997 for discussion of three modes of pulsar inner gap).

3. The low frequency waves can propagate near the neutron stars. A possible
reason for this may be that large radiation pressure may make particle's
density along the path of the emission to be substantially less dense than
that predicted (e.g. Sincell and Coppi 1996), and the plasma frequency should
be much lower if nonlinear effects are taken into account (Chian and Clemow
1975).

\subsection{The luminosity of the radio emission}

The efficiency of the ICS process is higher than that of the CR process, but
as the estimate bellow, incoherent radiation in ICS process is inadequate in
explaining pulsar radiation either. We can write the luminosity of the ICS
process as follows (see appendix):

\begin{equation}
L_{ics,incoh}=(1.5\times 10^{33} erg/s) \zeta B_{12}^3
h_3^2 P^{-4} \gamma_3 ^2 (\sigma/\sigma _{th})
\end{equation}

\noindent here $\zeta$ is in order of 1, $B_{12}=B/(10^{12} Gauss)$, $B$ is the
magnetic field near the surface of the neutron stars,
$h$ is the thickness of the gap, $h_3=h/(10^3 cm)$, $P$ is the rotational
period of the neutron star. $\sigma$ is the cross section of the inverse 
Compton process, and $\sigma _{th}$ is the Thomson cross section.

If we take $\sigma =\sigma _{th}/\gamma^2$ (see appendix), then

\begin{equation}
L_{ics,incoh}=(1.5 \times 10^{27}erg/s)\zeta B_{12}^3 h_3^2 P^{-4}
\end{equation}

When the radiation take place at a higher position: for example, $r=10R$,
using $n_{ph}=n_{ph,0}(R/r)^3$, in this case we have:

\begin{equation}
L_{ics,incoh}=(1.5 \times 10^{24}erg/s)\zeta B_{12}^3 h_3^2 P^{-4}
\end{equation}

The luminosity observed of radio pulsars can be written as (Sutherland 1979):

\begin{equation}
L=(3.3 \times 10^{25} ergs/s) S_{400} d^2
\end{equation}

\noindent where $S_{400}$ is the mean flux density in $mJy$ at $400 MHz$ and $d$ is the
pulsar distance in $kpc$. The ranges of $S_{400} d^2$ is from $\sim 10$ to $\sim
10^5 mJy-kpc^2$. This means that incoherent ICS radiation is inadequate in
explaining pulsar radiation. A coherent mechanism should be involved.

Coherent emission mechanisms may be classified as: (1).maser mechanisms;
(2).a reactive or hydrodynamic instability; or (3). to emission by bunches.
Theories for these coherent emission processes are not as well developed as
theories for incoherent emission processes (Melrose 1992).

The emission mechanism suggested in this paper are favorable to the coherent
emission by bunches. In the ICS process, the outgoing photons are produced
in a scattering process: the low frequency wave is scattered by particles
which is moving along a bunch magnetic field lines. In the process, the out
going photons produced by particles of a bunch are coherent; and for those
produced by particles in different sparking are not coherent. From this
point of view, we can take a calculation to fit the linear and circular
polarization observations of radio pulsars very well (Xu 1997; Qiao, Xu and Han 1997).
If the number of the particles in each sparking is $N_i$ ,the coherent
luminosity of the ICS process is: 

\begin{equation}
L_{ics,coh}=\sigma c n_{ph}\hbar \omega ' \sum \left( \frac{dN_i}
{dt}\right) ^2
\end{equation}

\noindent where $\frac{dN}{dt}=\sum \left( \frac{dN_i}{dt}\right)$. If $\frac{dN_i}{dt}
=\frac{dN_{i+1}}{dt}$, $\frac{dN}{dt}=n\left( \frac{dN_i}{dt}
\right)$, then the ratio of coherent luminosity to incoherent luminosity is:

\begin{equation}
\eta=L_{ics,coh}/L_{ics,incoh}=n \left(\frac{dN_i}{dt}\right)^2
/\frac{dN}{dt}=\frac{dN_i}{dt}
\end{equation}

Comparing Eq.(3) and (4), we find that $\frac{dN_i}{dt} \simeq 10^3 \sim 10^7$ is enough to produce observed radio emission.
From $n_0=n_{gj}\simeq \frac{\bf \Omega \cdot B}{2\pi ce}$ and $\frac{dN_i}{dt}
=2\chi \pi r_p^2 n_0 c$, we get

\begin{equation}
\frac{dN_i}{dt}=\frac{4\pi \chi B R^3}{ce P^2}=2.7\times 10^{30} \kappa
B_{12}P^{-2}
\end{equation}

\noindent where $r_p=R\theta_p=R \sqrt{{2\pi R} /Pc}$ is the radius of inner gap,
and $R=10^{6}cm$ is the radius of the neutron stars. 
For typical parameters and $\kappa \simeq 1$, $\frac{dN}{dt}$ is about $10
^{30}$, this means that only few percentage ($10^{-27}$ to $10^{-23}$) of
the coherent particles is enough to produce the observed radio emission.

For coherent curvature radiation, there is a fundamental weakness in
existing theoretical treatments which do not allow for any velocity
dispersion of the particles (Melrose 1992). In the mechanism discussed here
the weakness is much weaker. This is because for the observed emission 
(including frequency and phase) is determined by the frequency of the
incoming wave $\omega_0$, the energy of the particles (Lorentz factor
$\gamma$) and the incoming angle $\theta_i$, not only $\gamma$ (see Eq.(8)).

\subsection{The basic formulae for emission beams}

For most pulsars, $B \ll B_q=4.414\times 10^{13} Gauss$ at the points near or
far from the surface of neutron star. As Lorentz factor $\gamma \gg 1$,
$\beta =v/c\simeq 1$, $\gamma \hbar \omega _0\ll m_e c^2$ and $\theta '
\sim 0$, in an ICS process of the outgoing high energy particles with the 
low frequency wave photons for the outgoing photons (with the energy
of $\hbar \omega '$), we have (Qiao 1988a,b): 

\begin{equation}
\omega '=\gamma ^2\omega _0(1-\beta cos\theta _i)(1+\beta)
\simeq 2\gamma ^2\omega _0 (1-\beta cos\theta _i)
\end{equation}

\noindent here, $\theta _i (\theta ')$ is the incoming (outgoing) angle
between the direction of motion of a particle and the incoming (outgoing) photon, $m_e$
is the mass of electron, $\omega _0$ is the angular frequency of the low
frequency wave produced in the inner gap sparking. In the Lab frame, most of
the outgoing photons are emitted along the direction of motion of a particle
within a small beam which width is about $\gamma ^{-1}$. 

\begin{figure}
\centerline{\psfig{figure=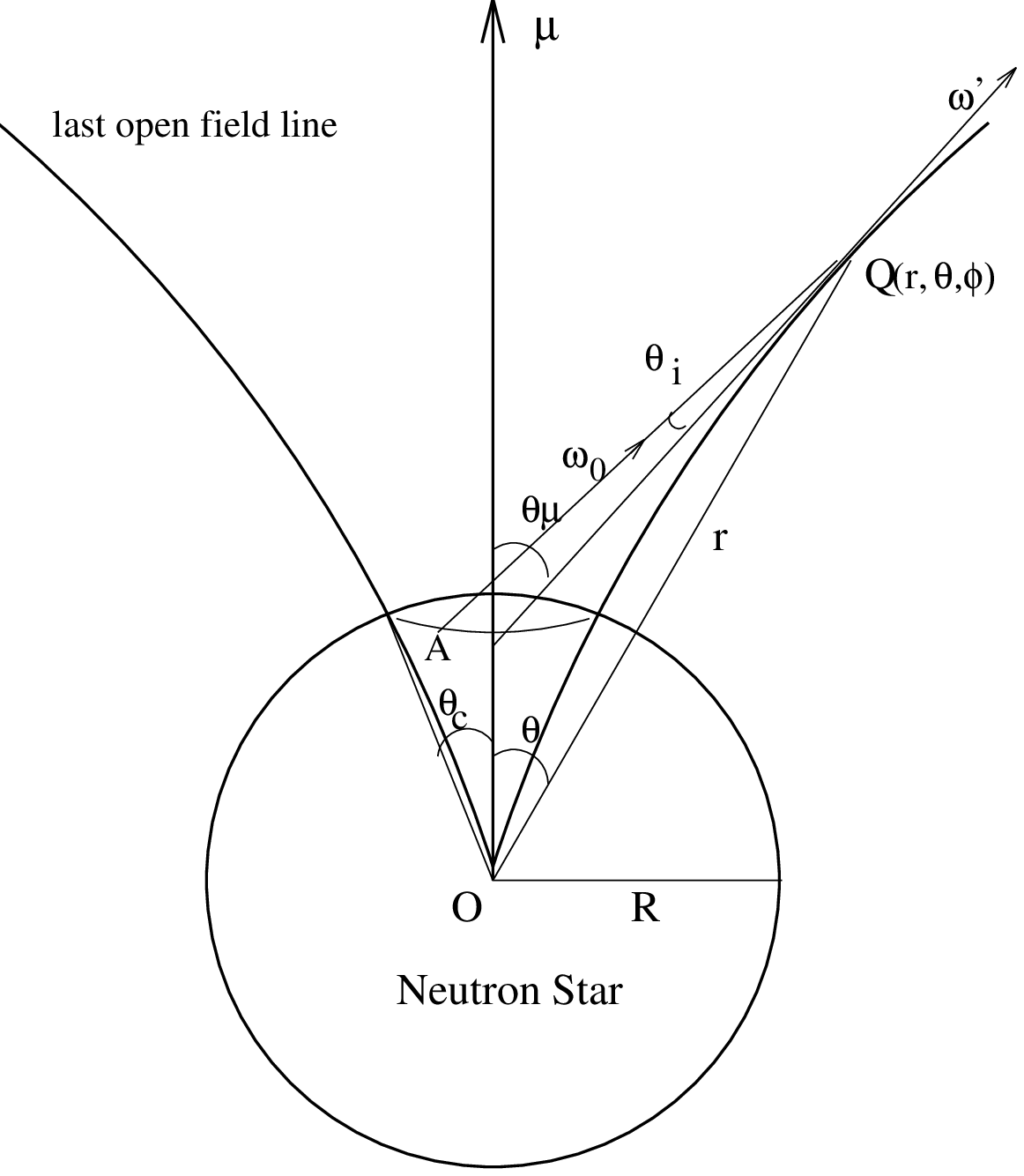,angle=270,height=6.5cm}}
      \caption{Geometry for the inverse Compton scattering process.
            Low frequency waves are scattered by particles at point 
            $Q(r, \theta, \phi)$. Dipole field line is assumed. }
         \label{FigGeo}
    \end{figure}

For a dipole magnetic field line, we have 

\begin{equation}
r=R_e sin^2\theta
\end{equation}

\noindent where $r$ is the distance between a point $Q$ and the center of the neutron
star, $R_e=\lambda R_c$ ($\lambda \geq 1$), $R_c={Pc/2\pi}$ is the radius of light cylinder,
$P$ is the pulse period of pulsar, $R$ is the radius of the neutron star, 
$\theta$ is the polar angle at point $Q(r,\theta,\phi)$ with respect to
the magnetic axis (see Fig.~\ref{FigGeo}), $\lambda (\geq 1)$ is a constant for a
dipole magnetic field line. For investigating the radiation, we only
consider so-called open field lines.

We consider the case that the low frequency wave is produced at the sparking
point $A$ near the boundary of the inner gap defined by the last open field
lines. At any scattering point $Q(r,\theta,\phi)$ in a field line with
$R_e$, the incident angle $\theta _i$ (see Fig.~\ref{FigGeo}) can be written as follows, 

\begin{equation}
cos\theta_i={{\bf B}\cdot {\bf AQ}}/\left( B\cdot AQ\right)
\end{equation}

\noindent here, ${\bf B}$ is the vector of magnetic field, ${\bf AQ}$ is the direction
vector of incoming low frequency wave.

In a right angle coordinate system a magnetic field can be written as: 

\begin{equation}
\left\{ 
\begin{array}{ll}
B_x & =\left( B_rsin\theta +B_\theta cos\theta \right) cos\phi -B_\phi
sin\phi  \\ 
B_y & =\left( B_rsin\theta +B_\theta cos\theta \right) sin\phi +B_\phi
cos\phi  \\ 
B_z & =B_rcos\theta -B_\theta sin\theta
\end{array}
\right.  
\end{equation}

In a spherical coordinate system the dipole magnetic field will have
components (Shitov,1985):

\begin{equation}
\left\{
\begin{array}{ll}
B_r &={\frac m{r^3}}2cos\theta\\
B_\theta &={\frac m{r^3}}sin\theta\\
B_\phi &=0
\end{array}
\right.
\end{equation}

\noindent and $B={\frac m{r^3}}\sqrt{1+3cos^2\theta }$. For the coordinate of point $A(x_a,y_a,z_a)$ and $Q(x_q,y_q,z_q)$, we have:

\begin{equation}
\left\{
\begin{array}{ll}
x_a &=R sin\theta _c cos\phi _c\\
y_a &=R sin\theta _c sin\phi _c\\
z_a &=R cos\theta _c
\end{array}
\right.
\end{equation}

\begin{equation}
\left\{
\begin{array}{ll}
x_q &=r\sin \theta \cos \phi\\
y_q &=r\sin \theta \sin \phi\\
z_q &=r\cos \theta
\end{array}
\right.
\end{equation}

Then vector ${\bf AQ}$ is:

\begin{equation}
{\bf AQ}=\{(x_q-x_a),(y_q-y_a),(z_q-z_a)\}
\end{equation}

\noindent where the azimuthal angle $\phi$ ranges from 0 to $2\pi$ for different
field lines which are symmetric to the magnetic axis, and $\theta _c=(2\pi
R/Pc)^{1/2}$ is the pole cap angle where the last open field line begins
(see RS75), the azimuthal angle of the inner gap boundary $\phi _c$ can be
from 0 to $2\pi$.

With Eqs(8) to (13) we can get $\theta _i$ easily:

\begin{equation}
cos\theta_i =M/N
\end{equation}

\noindent where
$M=2rcos \theta-R[3cos\theta
\sin\theta sin\theta_c cos(\phi-\phi_c)+(3cos^2 \theta-1) cos \theta_c]$, and
$N=(1+3 cos^2\theta)^{1/2}\{r^2+ R^2-2Rr[cos\theta cos \theta_c+sin\theta sin\theta_c cos(\phi-\phi_c)]\}^{1/2}$.

When the emission regions are far form the surface (that is for $r\gg R$,
and $\theta _c \approx 0$) and in the plane of a field line (that is $\phi=\phi_c$), we have:

\begin{equation}
cos\theta_i=\frac{2 cos\theta +(R/r) (1-3 cos^2 \theta)} 
{\sqrt{(1+3 cos^2\theta)[1-2(R/r) cos\theta+(R/r)^2]}}
\end{equation}

The angle between the radiation direction (in the direction of the magnetic
field) and the magnetic axis, $\theta _\mu$, has a simple relation with
$\theta$, 

\begin{equation}
ctg\theta _\mu =\frac{2ctg^2\theta -1}{3ctg\theta}
\end{equation}

The energy of high energy particles will be reduced when these particles come out along
with the field lines owing to scattering with thermal photons and low
frequency waves. It is assumed that: 

\begin{equation}
\gamma =\gamma _0\left[ 1-\xi {\left( r-R\right) /R_e}\right]
\end{equation}

\noindent where $\xi$ reflects the energy lose of the particles and different pulsars
have different $\gamma _0$ and $\xi$.

Using equations (8) to (19), we have a numerical relation between the
outgoing photon frequency $\omega'$ and the beam radius $\theta
_\mu$. In the coordinate system with magnetic axis as $z$ axis, we have
$\phi_\mu=\phi$. Finally, at any scattering point, we can get $\omega'$
and the radiation direction defined by $\theta_\mu$, $\phi_\mu$.

\subsection{Retardation and aberration effects}

Both observations (Rankin 1983a,b,1986,1990,1993; Lyne and Manchester,1988)
and calculations (this paper and Qiao et al. 1992) show that
the cone, ``inner'' cone and ``outer'' cone are emitted at different height
(see Fig.~\ref{FigHeight}), this makes the apparent beams move their positions relative to
each other. Two points are considered in this paper: First, it needs time
when the low frequency wave photons propagate to the points where they are
scattered by relativistic particles. For the scattering process taking place
at point $Q(r,\theta,\phi)$ (see Fig.~\ref{FigGeo}), the emission point of the low
frequency wave is not at point $A$, but at a point before that, as a result
the incoming angle $\theta _i$ will be changed. In other words, for the
scattering that takes place at this moment, the low frequency wave doesn't
come in from the present gap but the gap in an earlier position. The angle
difference is $\Delta \theta =\Delta t/P\times 360^{\circ }$, where $\Delta
t$ is the time for light to travel between the point where the low
frequency wave is emitted and the point where it is scattered. Secondly, the
core and two cones are emitted at different heights, hence a time delay
between them would change the apparent positions. Thus, we can get a new
$\theta_i$ this makes that the beams are asymmetric to the magnetic axis:
the central beam close to the trailing component of the cone, see Fig.~\ref{FigBeam2}.

\subsection{Basic results}

\begin{figure}[tbp]
\centerline{\psfig{figure=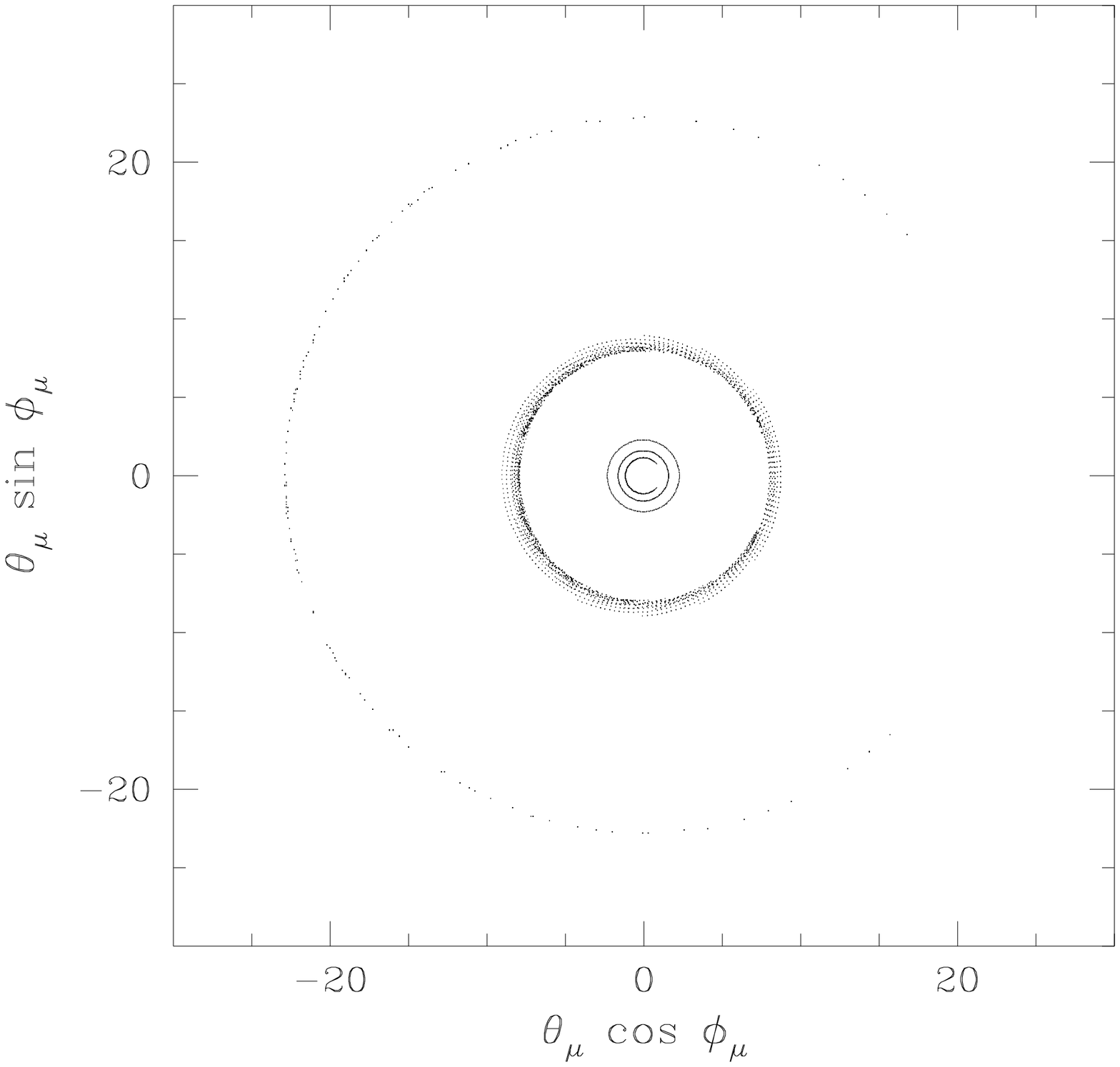,height=6.5cm}}
\centerline{\psfig{figure=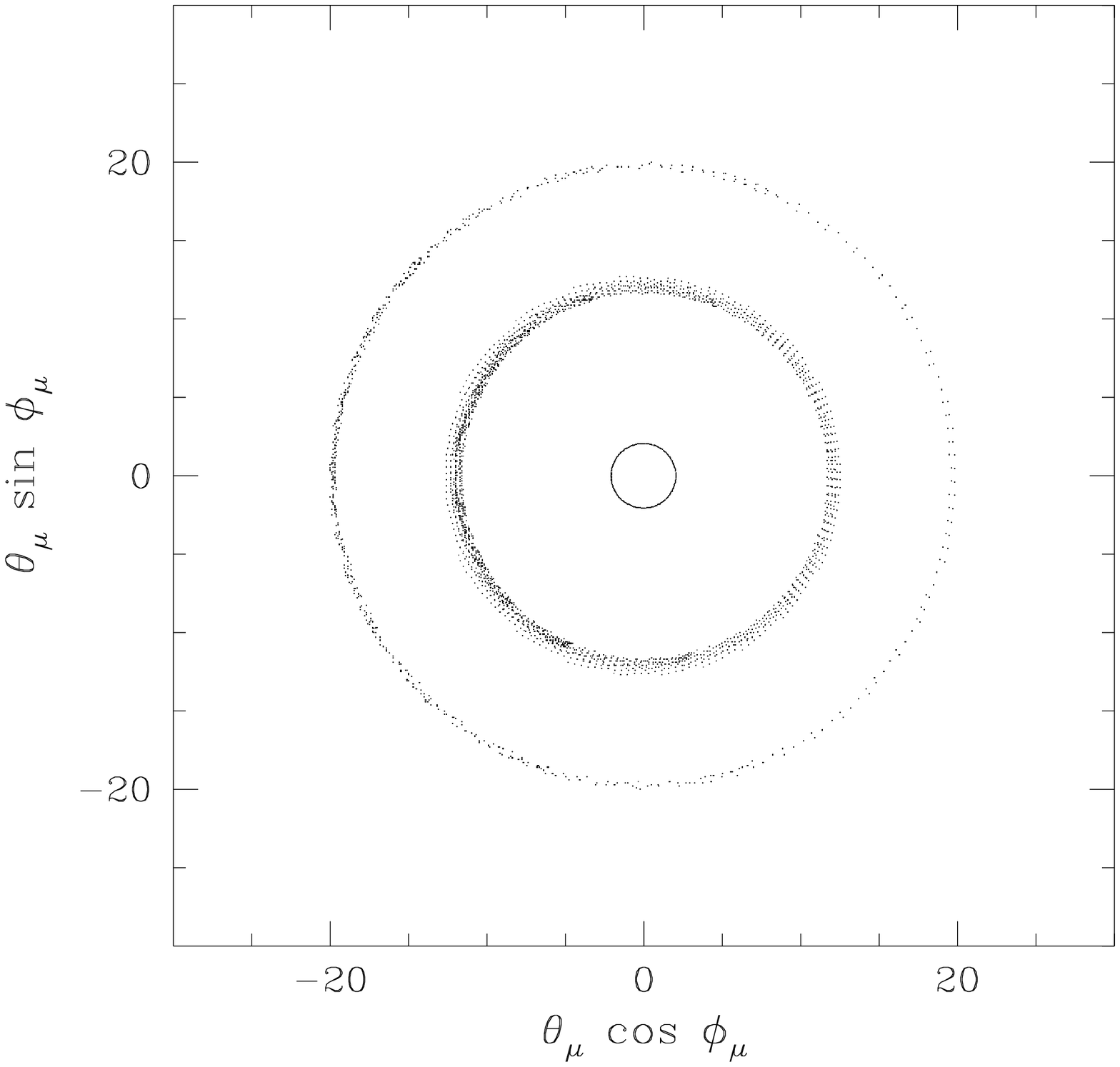,height=6.5cm}}
 \caption{Emission beams
for a pulsar with $P=1s$, $\alpha=45^{\circ}$, $\gamma_{0}=10^{3}$, $\omega_{0}=
10^{6} s^{-1}$, $\xi=1.0 \times 10^{1}$:
{\bf a}. at $350 MHz$ (upper panel); {\bf b}. at $500 MHz$ (lower panel).
The beams are slightly asymmetric to magnetic axis (which is perpendicular to 
the paper plane at the center of core). Retardation and aberration effects 
are weak for slow rotation pulsars.}
 \label{FigBeam1}
\end{figure}
\begin{figure}[tp]
\centerline{\psfig{figure=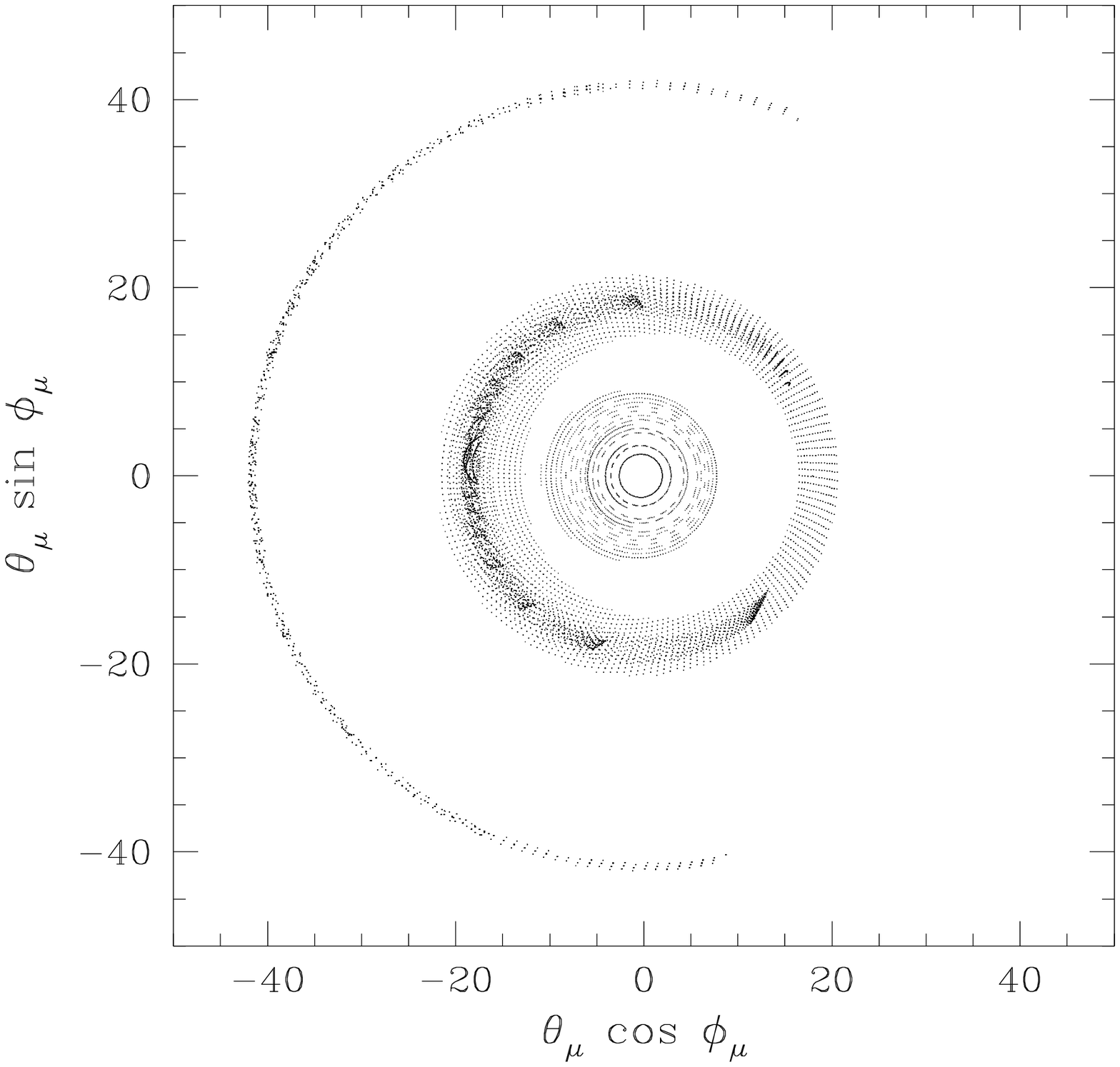,height=6.5cm}}
\centerline{\psfig{figure=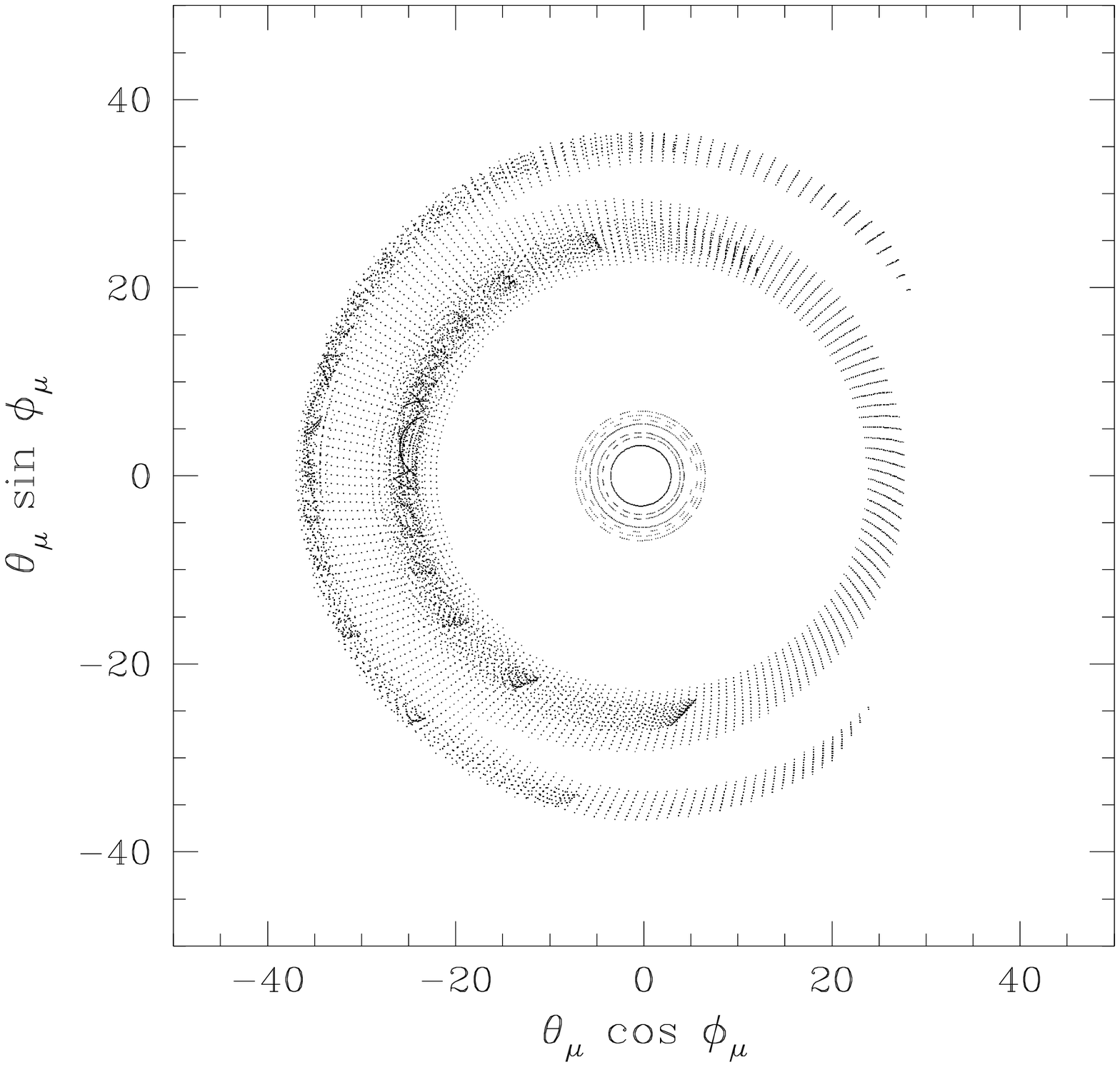,height=6.5cm}}
\caption{Emission beams with obvious retardation and aberration effects:
{\bf a}. at $350 MHz$ (upper panel); {\bf b}. at $500 MHz$ (lower panel).
$P=0.3s$, $\alpha=45^{\circ}$, $\gamma_{0}=5 \times 10^{2}$,
$\omega_{0}=10^{6} s^{-1}$, $\xi=5 \times 10^{-1}$.
The beams are asymmetric to magnetic axis.}
\label{FigBeam2}
\end{figure}
\begin{figure}[tp]
\centerline{\psfig{figure=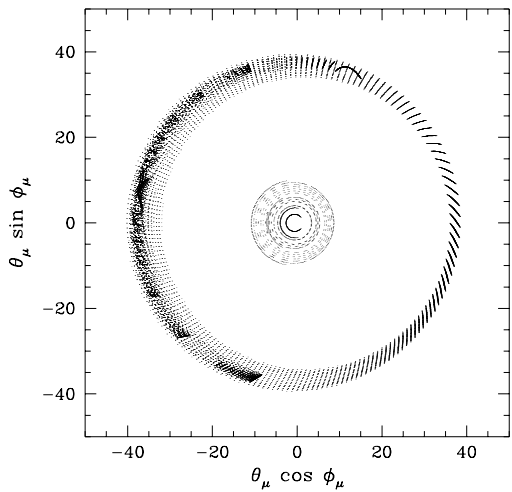,height=6.5cm}}
\caption{Emission beams at $500 MHz$ for a fast rotation pulsar with
$P=0.1s$, $\alpha=45^{\circ}$, $\gamma_{0}=3 \times 10^{2}$,
$\omega_{0}=10^{6} s^{-1}$, $\xi=1 \times 10^{-1}$.}
\label{FigBeam3}
\end{figure}
\begin{figure}[tpbh]
\centerline{\psfig{figure=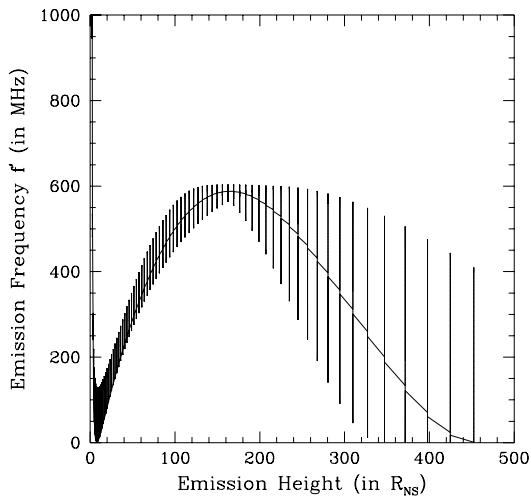,height=6.5cm}}
\caption{A relation between the frequency and the emission heights for a pulsar
with the same parameters of
Fig.~\ref{FigBeam1}. When observing at a frequency, one should find that the 
core, inner cone and outer cone are
emitted at different heights. The frequencies in the Figures in this paper only
have relative meaning.}
\label{FigHeight}
\end{figure}
\begin{figure}[ht]
\centerline{\psfig{figure=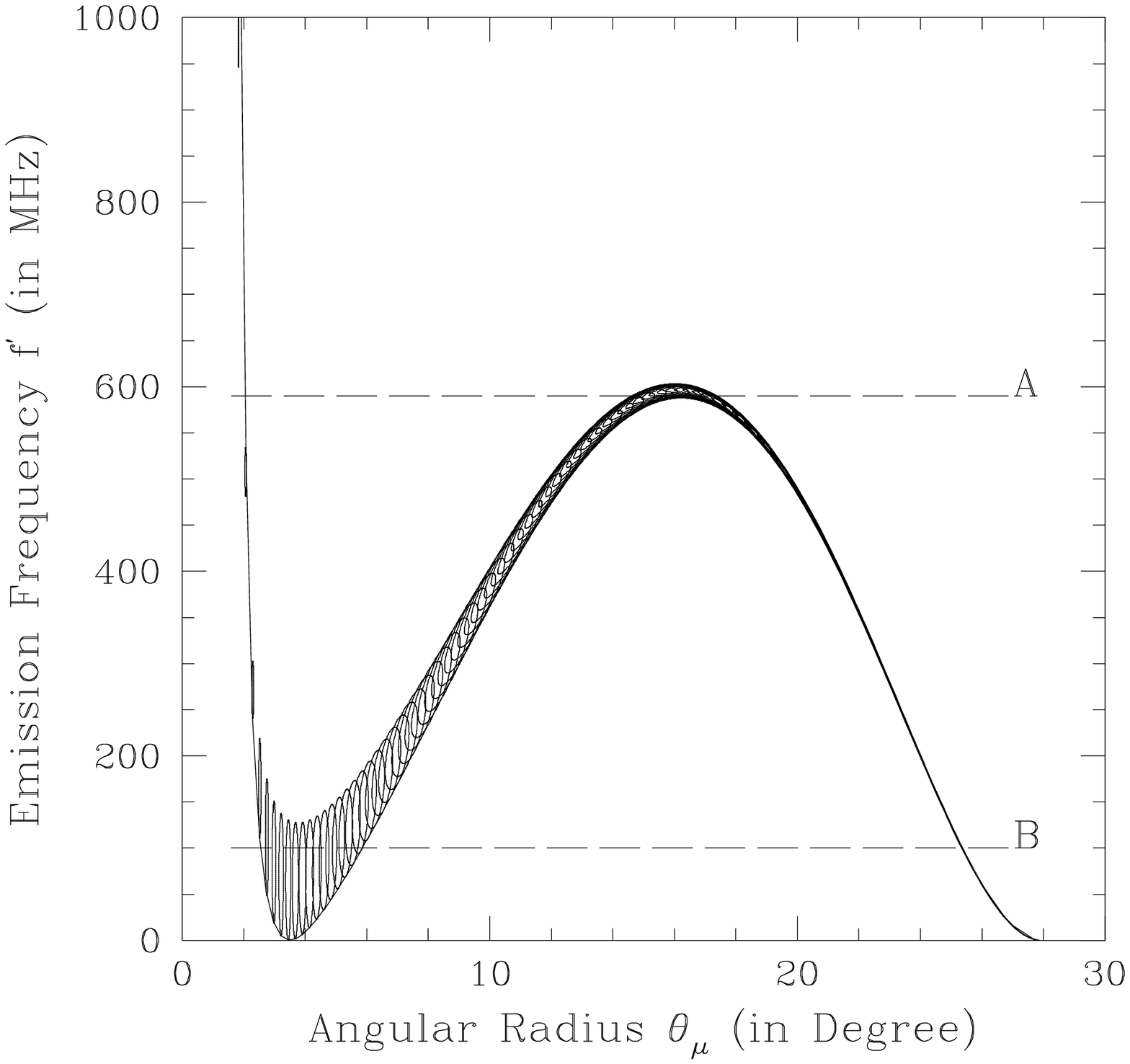,height=6.5cm}}
\centerline{\psfig{figure=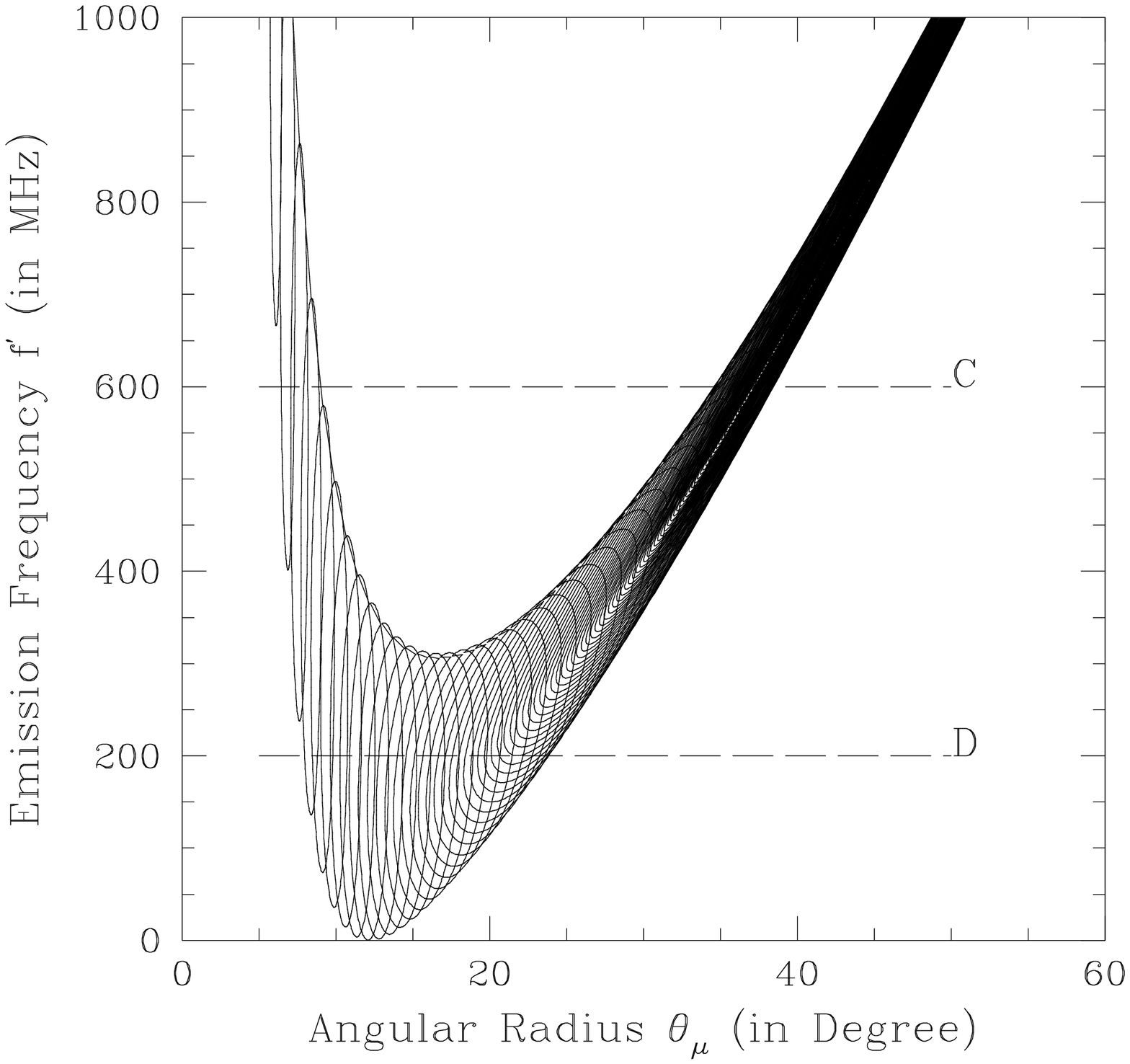,height=6.5cm}}
 \caption{Figure for $f'-\theta_{\mu}$ relation: {\bf a}.
with the same parameters of Fig.~\ref{FigBeam1} (upper panel);
{\bf b}. with the same parameters of Fig.~\ref{FigBeam3} (lower panel).}
 \label{FigFreq}
\end{figure}

The calculation results are shown in the Figures. According to our
calculated results, several main conclusions can be reached about the
emission regions:

\begin{enumerate}

\item The theoretical emission beams we get have two or three parts, including
``core'' and ``inner cone'' and/or an extra ``outer'' cone at a given
frequency. The angular radii of the beams are strongly related to the pulse
period $P$ and only $P$ (see Fig.~\ref{FigBeam1} and Fig.~\ref{FigBeam2}). This result is in good
agreement with the empirical figure of double-conal geometry for some
pulsars (Rankin 1993). The angular radius of central and conal beams are
consistent with the conclusion from observations (Rankin 1983,1993). Fig.~\ref{FigBeam1}
shows emission beams for a slow rotating pulsar with period $1.0s$, which
have one core and double cones. Fig.~\ref{FigBeam3} shows other emission beams for a fast
rotating pulsar with $P\sim 0.1s$, which only have one core and one cone and
it is very difficult to produce an ``outer'' cone for this kind of
short-period pulsars.

\item  These three different emission components are emitted at different
heights (Fig.~\ref{FigHeight}). The core emission is emitted at a place very close to the
surface of the neutron stars in a ``pencil'' beam and the ``inner'' cone is
emitted at a lower height along the same group of peripheral field lines
where the ``outer'' cone is produced. This is to say that radio emission
with the same frequency can be emitted at different heights along a bunch of
peripheral field lines, which is in agreement with the conclusion giving by
Rankin (1993).

\item Considering the two effects of retardation and aberration, the theoretical
beams for those pulsars with fast rotation will change their shapes. The
apparent position of the core will move to later longitudes with respect to
the position of the center of the cones.

\item In Fig.~\ref{FigFreq}, at relatively low emission frequencies, core and inner cone can
merge together (line B \& line D) And at high frequencies, inner cone and
outer cone, can merge together for slow rotating pulsars (line A). This can
be seen clearly in Fig.~\ref{FigBeam2}.

\item In some case (Fig.~\ref{FigBeam2} and Fig.~\ref{FigBeam3}), the leading part of outer cone can be
broken.
\end{enumerate}

\section{Discussion and conclusion}

\subsection{Shape of emission beams and pulse profiles}

The result in this paper shows that slow-rotation pulsars with long period
such as $0.5s-1.0s$ or longer, are ``double-conal pulsars'' with both inner
and outer cone beams. This is in agreement with the conclusion given by
Rankin (1993). Rankin's table (1993) for {\bf M} pulsars with five
components shows, most of these 19 pulsars have long pulse periods. In the
Fig.~\ref{FigFreq}a, the opening angle (angular radius) becomes larger
when the frequency increases (decreases) for the inner cone (outer cone).
Fast pulsars, i.e. $P<0.3s$, may only have a core and one inner conal
emission component (Fig.~\ref{FigBeam3} and Fig.~\ref{FigFreq}b). For this class
of pulsars, we suppose that observations tend to get triple ({\bf T})
profiles. In Fig.~\ref{FigFreq}b, as the frequency increases, the
opening angle for the inner cone always increases. In this case, at higher
frequency (from line D to line C), we will get a wider pulse profile (Qiao
1992) in contrast to slow pulsars. This can be seen for some pulsars, such
as PSR B1642-03 and PSR B1933+16 (Sieber et al. 1975).

We must mention that the theoretical angular radius of the outer conal beam
is larger than in previous calculations and the height of the outer conal
beam emission region is larger than that of Rankin's. This may be related to
the parameter $\xi $ in Eq.(5). A detailed calculation will show that the
controller $\xi $ is determined by the energy loss of particles, which
depends on the strength of the magnetic field and the thermal temperature at
the surface of neutron stars (Zhang te al. 1997).

\subsection{Does the ``inner'' cone radius increase as the observed
frequency increases?}

One result of this paper is that the inner cone radius increases as the
observed frequency increases. This result is supported by analysis of
observations. Wu et al.(1992)
present a method to deal with the structure of the mean pulse profiles of
pulsars. With that method and multi-frequency observational data, a diagram
of $f'-\theta_\mu$ was given, which is very similar to the
result of our calculation (see Fig.~\ref{FigFreq}) and Fig.2 of Qiao (1992).
Further analysis of
Rankin (1993) did not emphasize that the angular radius of the inner cone
$\rho_{inner}$ increases when the frequency increases. More analysis of this
is needed. A very direct method to check the result of this paper is
that: for pulsars with ``inner'' and ``outer'' cone (five components), the
pulse profiles would become a triple (``inner'' cone and ``outer'' cone get
together) with a smaller central component at very high frequency (Fig.~\ref{FigFreq},
line A); And also become a triple (the inner cone and core get together)
with stronger central component at low frequency (Fig.~\ref{FigFreq}, line B). This is in
agreement with the observations, {\em e.g.} Izvekova et al. (1989).

\subsection{conclusion}

Our result shows, ICS process is a possible radiation mechanism for radio
pulsars since it can produce the emission beams naturally and well
consistent with observations. Rankin (1993) showed that the angular radii of
core, ``inner'' cone and ``outer'' cone at a given frequency is a function
of P (and only P!). This is just the result of the calculations in this
paper. In agreement with the results given by Rankin (1993), we conclude
that those pulsars only with ``inner'' cone (core single and triple in
Rankin's classification) are generally faster, those with ``outer'' cone
(conal single and double) much slower, and the group of five-component ({\bf %
M}) pulsars falls in between the other two. This paper also supports that
the ``inner'' cone is emitted at a lower height along a same group of field
lines that produce the ``outer'' cone. The shapes of pulse profiles change
with frequencies in agreement with some kind of pulsars (Qiao 1992). The
retardation and aberration effects induce asymmetry and can be observable,
our result fits with observations (McCulloch 1992). These two effects may
also change the linear polarization position angle (Xu, et al. 1996).
The coherent emission of the ICS process suggested in
this paper is an efficient mechanism to produce observed luminosity, and
which is also a mechanism to produce observed polarization characters (Qiao
et al. 1997).

\acknowledgements {We are very grateful to Professor M.A. Ruderman for his
very impotent comments and suggestions. The authors thanks to Zhang B., Xu
R.X., Han J.L., Gao K.Y. and Liu J.F. for helpful discussions. This work is
partly supported by NSF of China, the Climbing Project-the National Key
Project for Fundamental Research of China, and the Project supported by
Doctoral Program Foundation of Institution of Higher Education in China.}

\appendix
{\flushleft {\bf Appendix}}

\section{The efficiency of Curvature Radiation (CR) and the ICS processes}

The energy loss of a particle through CR process is:

\begin{equation}
p_{cr}=\frac{2e^2c\gamma ^4}{3 \rho ^2}
\end{equation}

\noindent where $\gamma$ is the Lorentz factor of the particles, e is the electric
charge of the particles, c the light of speed, $\rho$ is the radius of
curvature of the magnetic field. The curvature radius $\rho$ for dipole
magnetic field is $\rho =\rho \sim \frac 43\left(rR_e\right) ^{1/2}$, $R_e =
\lambda R_c$, $R_c$ is the radius of the light cylinder, for last opening
field lines the $\rho$ is:

\begin{equation}
\rho =\rho \sim \frac 4 3\left( rR_c\right)^{1/2} \sim 10^8 cm P^{1/2}
\end{equation}

The energy loss of a particle through ICS is:

\begin{equation}
p_{ics}=\sigma cn_{ph0}\hbar \omega' 
\end{equation}

where

\begin{equation}
\omega' \simeq 2 \gamma^2 \omega _0(1-\beta \cos \theta_i)
\end{equation}

\noindent and $\sigma$ is the total cross-section of ICS, $n_{ph0}$ is the photon
number density near the surface, $\hbar \omega'$ is the
energy of the out going photons, near the surface, $\theta_i \simeq \pi
/2 $ and $\omega ' \simeq 2\gamma ^2\omega _0$ can be taking in
the estimate below, near the surface of the neutron stars, the photon number
density can be written as:

\begin{equation}
n_{ph0}=\frac s{c\hbar \omega_0}
\end{equation}

\begin{equation}
s=\frac c{4\pi}E^2
\end{equation}

\noindent where $E$ is the electric field in the gap, as the inner gap sparking, one
expects that the low frequency wave with electric field as the value of $E$.
In the limit $h\ll r_p$ ($h$ is the thickness of the gap, $r_p=R\theta_c$ is the
radius of the gap), we have (RS75):

\begin{equation}
E=2\frac{\Omega B h}c
\end{equation}

Where $B$ is the magnetic field near the surface of the neutron star. Substitute Eqs above to Eq.(A3), we have:

\begin{equation}
p_{ics}=\frac{2B^2\gamma ^2h^2\Omega ^2\sigma }{c\pi }
\end{equation}

So the ratio $\eta$ of the energy loss of these two processes near the
surface of the neuron star is:

\begin{equation}
\eta '=\frac{p_{cr}}{p_{ics}}=0.8\times 10^{-15}\gamma
_3^2P_{}^2B_{12}^{-2}h_3^{-2}\rho _8^{-2}\left( \frac \sigma {\sigma
_{th}}\right) ^{-1}
\end{equation}

If we use $\rho \sim 10^6cm$ (RS75), the ratio $\eta$ can be increase about 
$10^4$ times, but $\eta$ is still very small. This means that near the
surface of the neutron star the efficiency of CR compare which the ICS is
very low. Above the surface $\omega' \simeq \gamma ^2\omega _0$
can be taking, and

\begin{equation}
n_{ph}=n_{ph0}\left( \frac Rr\right)^3
\end{equation}

From the Eqs. above, we have:

\begin{equation}
\eta ' =\frac{p_{cr}}{p_{ics}}=\frac{3\pi c^2e^2\gamma ^2r^2
}{8R^3B^2h^2\Omega ^2R_c\sigma }
=1.9\times 10^{-5}\gamma _3^4 P r_8^2B_{12}^{-2}h_3^{-2}
\end{equation}

\noindent where $\sigma \sim \gamma^2 \sigma_{th}$ (see below). Even if the typical radiation height is at $100km$ in Eq.(A11), $\eta '$ is
still small. This means that the efficiency of the ICS process is higher
than that of the CR process, but as the estimate bellow, the
incoherent radiation of ICS process is inadequate in explaining pulsar
radiation either.
\bigskip

\section{Average number densities of the out flow}

The energy flux carried by relativistic positrons into the magnetosphere
above two polar gaps are (RS75):

\begin{equation}
\left( \frac{dE}{dt}\right) _{cur}=2JdV\pi r_p^2
\end{equation}

\noindent where the current form in the magnetosphere is taken as (Sutherland, 1979):

\begin{equation}
J=J_{rot}$ + $J_{\parallel }
\end{equation}

\noindent here $J_{rot}=\rho (\Omega \times r)$, the current associated with the
corotating plasma of charge density is (Goldreich and Julian 1969):

\begin{equation}
\rho _{gj}=\frac{\bf \nabla \cdot E}{4\pi }=-\frac{\bf \Omega \cdot B}{2\pi c}%
(1-\Omega ^2r^2\sin {}^2\theta /c^2)^{-1}
\end{equation}

If the magnetospere is charge separated then the number density of the
particles of charge is

\begin{equation}
n_0=n_{gj}\simeq \frac{\bf \Omega \cdot B}{2\pi ce}
\end{equation}

The second current corresponds to streaming of charges along the magnetic
field lines: $j_{\parallel}=kB$, $k$ must be constant along a given field
line. This current only exist on the open field lines. The exact division
between the open and closed field lines cannot be determined precisely. We
may use the vacuum dipole field geometry to locate approximately the
division on the neutron star and use the Eq.(B4).

If the potential in the inner gap has the maximum value (RS75):

\begin{equation}
dV=dV_{\max }=\frac{\Omega B r_p^2}{2c}
\end{equation}

We get

\begin{equation}
\left( \frac{dE}{dt}\right) _{cur}\simeq \left| \frac
{dE}{dt}_{dip}\right| =\frac{2\Omega ^4R^6B^2}{3c^3}
\end{equation}

This means that the current flow from the gaps have enough braking torque on
the spinning star and loss all the rotational energy (Sutherland 1979). This
is reasonable, because $n_{gj}=n_{\div }-n_{-}$, Even if in the charge
separated magnetosphere, n$_0$ can be much larger than the GJ density
$n_{gj}$. Especially, in the place near the stellar surface. Near stellar
surface, in the balance between the gravitational force and kinetic energy
of the particles lets a thin atmosphere existence. We will estimate the
luminosity of the ICS process using Eq.(B4).

\bigskip
\section{Luminosity in incoherent ICS processes}

The incoherent Luminosity by the ICS process near the
surface of the neutron star is:

\begin{equation}
L_{ics,incoh}=\sigma cn_{ph}\hbar \omega '\frac{dN}{dt}
\end{equation}

\begin{equation}
\frac{dN}{dt}=2\chi \pi r_p^2 n_0 c
\end{equation}

Substitute Eq.(A4)(B4) and (C2) to Eq.(C1), we can get the luminosities
as follows:

\begin{equation}
L_{ics,incoh}=(1.5\times 10^{33}erg/s)\zeta B_{12}^3\gamma _3^2 h_3^2
P^{-4} \left(\frac \sigma {\sigma _{th}}\right)
\end{equation}

\noindent where $\zeta$ is in order of 1, $\sigma_{th}$ is the Thomson cross section.

The ICS cross section $\sigma$ for the low frequency discussed in this paper
is (Qiao et al. 1986, Xia et al 1986) :

\begin{equation}
\begin{array}{ll}
\sigma _T(1)& =\sigma (1\rightarrow 1 ')+\sigma (1\rightarrow
2 ') \\
 & =\sigma _{th}\left\{ \sin^2\theta +\frac 12\cos^2\theta \left[
(\frac \omega {\omega +\omega _B})^2+(\frac \omega {\omega -\omega
_B})^2\right] \right\} ^R
\end{array}
\end{equation}

\begin{equation}
\begin{array}{ll}
\sigma _T(2)& =\sigma (2\rightarrow 1 ')+\sigma (2\rightarrow
2 ') \\
 & =\frac 12\sigma _{th}\left[ (\frac \omega {\omega +\omega _B})^2+(\frac
\omega {\omega -\omega _B})^2\right] ^R
\end{array}
\end{equation}

Here index ``$R$'' presents the parameters in the electron rest frame. In our case $\omega ^R\ll \omega _B$, so $\sigma _T(1)\gg \sigma _T(2)$, and

\begin{equation}
\sigma _T(1)=\frac{\sin^2\theta _i}{\gamma ^2\left( 1-\beta \cos \theta
_i\right) }\sigma_{th}
\end{equation}

\noindent where $\theta_i$ is the angle between the direction of the in coming photon
and the magnetic field. As $\theta_i \simeq \pi /2$,

\begin{equation}
\sigma _T(1)\simeq \gamma ^{-2}\sigma _{th}
\end{equation}

\noindent or 

\begin{equation}
\sigma_T(1) \omega ' \sim \sigma_{th} \omega_0
\end{equation}

At the place $r\gg R$, we can use $n\simeq n_0\left( \frac Rr\right) ^3$, if
the radiation region is at $r=100km$, so the luminosity in Eq.(C3) will be
down 9 magnitude. The luminosity observed of radio pulsars can be written as
(Sutherland,1979):

\begin{equation}
L=(3.3\times 10^{25}ergs/s)S_{400}d^2
\end{equation}

The ranges of $S_{400}d^2$ is from $\sim 10$ to $\sim 10^5 mJy-kpc^2$, which
means that incoherent ICS radiation is inadequate in explaining pulsar
radiation.

\bigskip
\section{Coherent emission}

See the text.

\end{document}